\documentclass[aps,prl,twocolumn]{revtex4}
\usepackage{graphicx}
\usepackage{bm}

\renewcommand{\v}[1]{{\bf #1}}

\def\be{\begin{eqnarray}}
\def\ee{\end{eqnarray}}
\newcommand{\nn}{\nonumber\\}

\newcommand{\Eq}[1]{Eq.~(\ref{#1})}

\newcommand{\ra}{\rightarrow}

\begin{document}

%\preprint{\sf Version 1 (\today)}
\title{New family of models for incompressible quantum liquids in  $d\ge 2$}
\author{Chyh-Hong Chern$^1$}
\email{chern@issp.u-tokyo.ac.jp}
\author{Dung-Hai Lee$^2$}
\affiliation{$^1$ERATO-SSS, Department of Applied Physics,
University of Tokyo, Tokyo 113-8656, Japan \\ $^2$Department of
Physics, University of California at Berkeley, Berkeley CA 94720,
U.S.A.}

\begin{abstract}
Through Haldane's construction, the fractional quantum Hall states
on a two-sphere was shown to be the ground states of {\it
one-dimensional} SU(2) spin Hamiltonians. In this Letter we
generalize this construction to obtain a new class of SU(N) spin
Hamiltonians. These Hamiltonians describes center-of-mass-position
conserving pair hopping fermions in space dimension $d\ge 2$.
\end{abstract}

%\pacs{Valid PACS appear here}
\maketitle

Gapped quantum many-body systems are stable states of matter because
they are robust against weak perturbations.  For fermions, the most
common gapped system is the ``band insulator'' where an energy gap
in the dispersion relation separates filled and empty
single-particle states. Here the insulating behavior is caused by
the Pauli exclusion principle. We refer to a system as a ``many-body
insulator'' if its energy gap is caused by many-body interaction
rather than one-particle dispersion relation. It is widely believed
that such a many-body gap can exist when the occupation number
(i.e.,the averaged number of particle per spin per unit cell) is a
fraction. The Mott insulator, where the energy gap is due to the
inter-particle repulsion, is an example of many-body insulator.

Contrary to the common belief, it is very difficult to find true
many-body insulators. In most cases an energy gap at fractional
occupation number is  accompanied by the breaking of translation
symmetry. After symmetry breaking, the  unit cell is enlarged such
that the new occupation number is an integer. In a recent work
Oshikawa argued that the existence of energy gap at fractional
occupation requires the ground state to be
degenerate\cite{oshikawa}. It happens that under usual circumstances
such degeneracy is caused by translation symmetry breaking.

The fractional quantum Hall state is a quantum liquid (hence no
breaking of translation symmetry) with an energy gap. On the
surface it is not clear what does it have to do with the many-body
insulator discussed above. In
Ref.\cite{Lee2004PRL,Seidel2005,Seidel2006} it is shown that, when
placed on a torus, both abelian and non-abelian quantum Hall
liquids can be mapped to many-body insulators on a {\it one
dimensional} ring of lattice sites. When both dimensions of the
torus are much bigger than the magnetic length, the energy gap is
generated by a long range center-of-mass-position conserving
hopping, rather than density-density interaction.

Partly motivated by Anderson's proposal of spin liquid\cite{rvb},
the question of whether a many-body insulator can exist without
symmetry breaking in spatial dimension $d\ge 2$ has attracted a
lot of interests. In this Letter we give this question an
affirmative answer by explicitly constructing a new class of
solvable lattice models that exhibit incompressible quantum liquid
ground states. Since our construction is a generalization of
Haldane's work on the pseudopotential Hamiltonian for the
fractional quantum Hall effect\cite{Haldane1983PRL}, we shall
begin by briefly review it.

If a magnetic monopole of strength $2S$ ($S$ is a multiple of $1/2$)
is placed at the center of a two-sphere, the kinetic energy spectrum
of a particle confined to move on the sphere is given by
$E_{k}\propto (S+k)(S+k+1)$, where $k$ is an non-negative integer.
The $k$th ``Landau level'' is $2(S+k)+1$-fold degenerate. The
degeneracy of the lowest Landau level, $k=0$, is exactly the
dimension of a spin-$S$ SU(2) multiplet. Thus the Hilbert space of
$N$ spin polarized electrons in the lowest Landau level is the same
as the exchange-antisymmetric sub-Hilbert space of $N$ such SU(2)
spins. Haldane's pseudopotential Hamiltonian (which has the
spherical version of Laughlin's $\nu=1/m$ wavefunction as the ground
state) is given by
\begin{eqnarray}
H=\frac{1}{2}\sum_{i\ne j}\sum_{q=1,odd}^{m-2} \kappa_q
~P^{2S-q}_{ij}.\label{su2}
\end{eqnarray}
Here $i,j=1,..,N$ are the spin labels, $P_{ij}^{2S-q}$ projects the
product states of spin $i$ and $j$ onto the total spin $2S-q$
multiplet, and $\kappa_q>0$ are parameters (as a result \Eq{su2} is
positive-definite). In Eq.(\ref{su2}) the $q$-sum is restricted to
odd integers because the restriction of the Hilbert space to the
total antisymmetric subspace. It can be shown that when
$N-1=2S/m\equiv p$ the Hamiltonian in Eq.(\ref{su2}) has an unique
ground state described by the following spin coherent-state
wavefunction(the spherical version of Laughlin's wavefunction)
\begin{eqnarray} \Psi=\left|
\begin{array}{ccccc} u_1^p & u_1^{p-1}v_1 & . & . & v_1^p \\
 u_2^p & u_2^{p-1}v_2 & . & . & v_2^p \\. & . & . & . & .
\\ . & . & . & . & . \\ u_N^p & u_N^{p-1}v_N & . & . & v_N^p
\end{array} \right|^m \label{su2wf}
\end{eqnarray}
Upon expanding the determinant, Eq.(\ref{su2wf}) can be written as
a linear combination of $\prod_{j=1}^N\sqrt{(mp)!/n_j!k_j!}
u_j^{n_j} v_j^{k_j}$ where $n_j+k_j=mp$. Each term in this linear
combination is a direct product of $N$, spin-$S$, states. An
important property of the wavefunction in Eq.(\ref{su2wf}) is that
the highest total spin for any pair is $2S-m$. Consequently
Eq.(\ref{su2wf}) is a zero energy state of \Eq{su2}. Moreover, if
we view the $2S+1$ different $S_z$ states as the $2S+1$ local
orbitals of a one-dimensional lattice, and write
$P^{2S-q}_{ij}=\sum_{l=-(2S-q)}^{2S-q}\sum_{m_2=-S}^S\sum_{m_1=-S}^S
C^{2S-q,l}_{S,m_1,S,l-m_1}C^{2S-q,l}_{S,m_2,S,l-m_2}|m_2,l-m_2><m_1,l-m_1|$,
Eq.(\ref{su2}) becomes a center-of-mass conserving pair hopping
Hamiltonian.\cite{Lee2004PRL,Seidel2005} In the above
$C^{2S-q,l}_{S,m_1,S,l-m_1}$ is the SU(2) Clebsch-Gordon
coefficient. The role of center-of-mass position conservation in
producing true many-body insulators was discussed in
Ref.\cite{Lee2004PRL,Seidel2005}.

\begin{figure}[htb]
\centerline{\includegraphics[angle=0,scale=0.2]{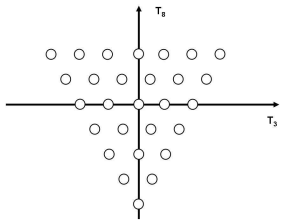}}
\caption{The weight space of $(6,0)$.} \label{Fig:weight_su3_6_0}
\end{figure}

In the following we generalize Haldane's construction to SU(3)
spins. The reason for doing so is SU(3), a rank two Lie group, has
multiplets isomorphic to {\it two dimensional} lattices. Hence it
allows us the possibility of constructing Hamiltonians for
many-body insulator in $d=2$. The irreducible representations of
SU(3) are labelled by two integers $(p,q)$. In the following we
shall focus on the the multiplets $(k,0)$. The reason is because
these multiplets are the only ones whose weight space is an array
of {\it non-degenerate}, i.e., non-duplicated, points (see
Fig.(\ref{Fig:weight_su3_6_0})). In the following we consider
SU(3) spins each in the $(mp,0)$ representation. The dimension of
the $(mp,0)$ representation is $d(mp)=(mp+1)(mp+2)/2$. For reason
that shall become clear later, we shall choose the number of spins
so that $N=d(p)=(p+1)(p+2)/2$. Under such condition the filling
factor, $f=d(p)/d(mp)$, is $1/m^2$ in the thermodynamic
($p\rightarrow\infty$) limit. As earlier, we will constrain the
N-spin Hilbert space to be exchange-antisymmetric to mimic the
fermion statistics.

The spin Hamiltonian we construct is a generalization of \Eq{su2},
and is given by
\begin{eqnarray}
H=\frac{1}{2}\sum_{i\ne j}\sum_{q=1, \ \text{odd}}^{q\leq m\!-\!2}
\kappa_q~P_{ij}^{(2mp\!-\!2q,q)}. \label{su3}
\end{eqnarray}
Here the operator $P_{ij}^{(2mp\!-\!2q,q)}$ operates on the direct
product states of two spins $i$ and $j$, and projects them onto
the $(2mp-2q,q)$ multiplet, and $\kappa_q>0$. For simplicity in
the rest of the Letter we shall set $m=3$ and $p=$ odd integer. In
this case Eq.(\ref{su3}) becomes
\begin{eqnarray}
H=\frac{\kappa_1}{2} \sum_{i\ne j}
P_{ij}^{(6p\!-\!2,1)}.\label{su31}
\end{eqnarray}

We shall spend much of the rest of the Letter to prove that
Eq.(\ref{su31}) has a unique singlet ground state described by the
following SU(3) coherent-state wavefunction
\begin{eqnarray}
\Psi_3 = \left|
\begin{array}{ccccc} u_1^p & u_1^{p-1}v_1 & . & . & w_1^p \\ u_2^p
& u_2^{p-1}v_2 & . & . & w_2^p \\ . & . & . & . & .
\\ . & . & . & . & . \\ u_N^p & u_N^{p-1}v_N & . & . & w_N^p
\end{array} \right|^3 \label{su3wf}
\end{eqnarray}
First we prove Eq.(\ref{su3wf}) is a ground state. Let us focus on
the dependence of \Eq{su3wf} on the variables of any chosen pair
of spin $i$ and $j$. For each of the Slater determinant in
\Eq{su3wf} the highest total SU(3) weight of these two spins is
$(2p-2,1)$ because of antisymmetry. As a result when we multiply
three  determinant together the highest total SU(3) weight for
spin $i$ and $j$ is $(6p-6,3)$. Consequently Eq.(\ref{su3wf}) is
an zero-energy eigenstate of the positive-definite Hamiltonian in
Eq.(\ref{su31}). Thus we have found a ground state.

As to the uniqueness let us start with the simplest case of $p=1$
where the single spin Hilbert space is $d(3)=10$ dimensional and
there are $d(1)=3$ spins. The most general many-body wave function
is given by
\begin{eqnarray}
\chi=\sum_{\{\alpha_{j},\beta_j\gamma_j=1\}}^3C(\alpha_{1},\alpha_{2},\alpha_{3};\beta_{1},\beta_{2},\beta_{3};\gamma_{1},\gamma_{2},\gamma_{3})\nonumber
\\\times
\phi_1^{\alpha_{1}}\phi_1^{\alpha_{2}}\phi_1^{\alpha_{3}}\phi_2^{\beta_{1}}\phi_2^{\beta_{2}}\phi_2^{\beta_{3}}\phi_3^{\gamma_{1}}\phi_3^{\gamma_{2}}
\phi_3^{\gamma_{3}},\label{chi}
\end{eqnarray}
where $\phi_j^{1,2,3}=u_j,v_j,w_j$. The requirement that this
wavefunction lies in the direct product of $(3,0)\otimes
(3,0)\otimes (3,0)$ demands the coefficient $C$ to be invariant
when the three indices of any chosen particle are permuted. In
addition, the antisymmetry constraint restricts $C$ to change sign
upon the exchange of particle labels.  Next, let us pick any pair
of spins $i$ and $j$ and examine the dependence of \Eq{chi} on
their variables.  The possible total SU(3) weights of these two
spins that are consistent with antisymmetry are given by
$(3,0)\otimes(3,0)=(4,1)\oplus(0,3)$. In order for the
wavefunction to be annihilated by \Eq{su31}, it must lie entirely
in $(0,3)$. Since there are only three particles it is not hard to
show that there is a unique $C$ so that the above condition holds
for all pair $(i,j)$: \be
C(\alpha_{1},\alpha_{2},\alpha_{3};\beta_{1},\beta_{2},\beta_{3};\gamma_{1},\gamma_{2},\gamma_{3})\nonumber
\propto\epsilon_{\alpha_{1}\beta_{1}\gamma_{1}}\epsilon_{\alpha_{2}\beta_{2}\gamma_{2}}\epsilon_{\alpha_{3}\beta_{3}\gamma_{3}}.\ee
Substitute the above equation into \Eq{chi} gives $\chi\sim
\Psi_3$.

Now, we consider $p=3$, where the single-particle Hilbert space is
55 dimensional and and there are $d(3)=10$ particles. Analogous to
\Eq{chi} the most general 10-particle wavefunction is given by
\begin{eqnarray}
\chi=\sum_{\{\alpha_{jn}=1\}}^3\!\! C(\!\{\alpha_{jn}\!\})\!
\prod_{j=1}^{10}\prod_{n=1}^9\phi_j^{\alpha_{jn}}. \label{Phi}
\end{eqnarray}
Here $j$ is the particle label, $n=1,...,9$ labels the nine
fundamental SU(3) spinors $(1,0)$ that make up $(9,0)$. The symmetry
properties of $C$ are the same as before. In this case the total
SU(3) weight of any two spins that are consistent with antisymmetry
are given $(9,0)\otimes (9,0)=(16,1)\oplus (12,3)\oplus (8,5)\oplus
(4,7)\oplus (0,9)$. The condition of being annihilated by the
Hamiltonian requires the ground state wavefunction to lie entirely
in $(12,3)\oplus (8,5)\oplus (4,7)\oplus (0,9)$, i.e.,
\begin{eqnarray}
(P_{ij}^{(12,3)}+P_{ij}^{(8,5)}+P_{ij}^{(4,7)}+P_{ij}^{(0,9)})\chi=\chi,
~~\forall ~(i,j). \label{su34}
\end{eqnarray}
\Eq{su34} implies that among the 9 indices spin $i$ and $j$ each
possesses, there must be at least 3 pairs (a pair contains one index
from each particle) such that $C\rightarrow -C$ upon exchanging the
indices within each pair. In addition to be consistent with exchange
antisymmetry, the number of such pairs also must be odd. Since $C$
is invariant when the nine indices of any particle are permuted, we
can perform permutations so that in each triplet
$(\alpha_{i1},\alpha_{i2},\alpha_{i3})$,
$(\alpha_{i4},\alpha_{i5},\alpha_{i6})$,$(\alpha_{i7},\alpha_{i8},\alpha_{i9})$
of particle $i$ and $(\alpha_{j1},\alpha_{j2},\alpha_{j3})$,
$(\alpha_{j4},\alpha_{j5},\alpha_{j6})$,$(\alpha_{j7},\alpha_{j8},\alpha_{j9})$
of particle $j$ there is an {\it odd number} of antisymmetric
indices. Under such circumstance $C$ changes sign upon {\it
independent} exchanges of the triplets, i.e., \be C\rightarrow
-C~~\text{upon}&&(\alpha_{i1},\alpha_{i2},\alpha_{i3})\leftrightarrow
(\alpha_{j1},\alpha_{j2},\alpha_{j3}) \nn
&&(\alpha_{i4},\alpha_{i5},\alpha_{i6})\leftrightarrow
(\alpha_{j4},\alpha_{j5},\alpha_{j6}) \nn
&&(\alpha_{i7},\alpha_{i8},\alpha_{i9})\leftrightarrow
(\alpha_{j7},\alpha_{j8},\alpha_{j9}).\label{exa3}\ee In \Eq{exa3}
$(...)\leftrightarrow (...)$ denotes the exchange of whole group of
indices. If \Eq{exa3} can be made true {\it simultaneously} for all
pairs $i$ and $j$ then $\Psi_3$ is the unique solution of \Eq{su34}.
This is proven as follows.

Let us focus on the dependence of $C$ on the first index triplet
of all particle. For each triplet of indices, say
$(\alpha_{i1},\alpha_{i2},\alpha_{i3})$, there are
$(3+2)!/(3!2!)=d(3)=10$ inequivalent combinations. We can
interpret each combination as a single-particle quantum state and
$C$ as the wavefunction for 10 particles to occupy these states.
The first line of \Eq{exa3} allows us to interpret $C$ as the
wavefunction for fermions. For $N=d(3)=10$, i.e., when the fermion
number is the same as the number of single-particle state, there
is an unique wavefunction satisfying the antisymmetric
requirement, namely,
\begin{eqnarray}
C(\{\alpha_{i1},\alpha_{i2},\alpha_{i3}\}...)\sim
\epsilon\{(\alpha_{i1}\alpha_{i2}\alpha_{i3})\}\label{c1}
\end{eqnarray}
where $\epsilon\{...\}$ is the rank 10 total antisymmetric tensor
with respect to the exchange of index-triplets. Similar argument can
be made to $\{(\alpha_{i4},\alpha_{i5},\alpha_{i6})\}$ and
$\{(\alpha_{i7},\alpha_{i8},\alpha_{i9})\}$, and lead to \be
&&C(\{\alpha_{i1},...,\alpha_{i9}\})\sim
\epsilon\{(\alpha_{i1}\alpha_{i2}\alpha_{i3})\}
\epsilon\{(\alpha_{i4}\alpha_{i5}\alpha_{i6})\}
\nn&\times&\epsilon\{(\alpha_{i7}\alpha_{i8}\alpha_{i9})\}.\label{c2}\ee
Substitute \Eq{c2} into \Eq{Phi} we obtain $\chi\sim \Psi_3$.

Now, we shall prove that Eq.(\ref{exa3}) can indeed be made true
for all pairs $i$ and $j$ simultaneously. Let us assume that there
exists a ground state solution whose $C$ does not satisfy
Eq.(\ref{exa3}) for pair $(k,l)$. This means there must be at
least one triplet exchange, let say
$\{\alpha_{k1},\alpha_{k2},\alpha_{k3}\}\leftrightarrow
\{\alpha_{l1},\alpha_{l2},\alpha_{l3}\}$, for which $C$ does not
transform according to \Eq{exa3}. However, since the wavefunction
still has to satisfy \Eq{su34} for $(k,l)$, we should be able to
write $C=C_3+C_5+C_7+C_9$, where $C_q$ is the component of $C$
that is odd with respect to exchange of {\it exactly} $q$ pair of
indices between particle $k$ and $l$ and even with respect to the
exchange of the rest. Now let us consider the effect of
$\{\alpha_{k1},\alpha_{k2},\alpha_{k3}\}\leftrightarrow
\{\alpha_{l1},\alpha_{l2},\alpha_{l3}\}$ on $C$.  Under such
operation $C_q$ can either change sign or stay invariant depending
on whether an odd or even number (out of $q$) antisymmetric
indices are contained in the specified triplets. In other words
upon $\{\alpha_{k1},\alpha_{k2},\alpha_{k3}\}\leftrightarrow
\{\alpha_{l1},\alpha_{l2},\alpha_{l3}\}$ we have \be C\rightarrow
\eta_3 C_3+\eta_5 C_5+\eta_7 C_7+\eta_9 C_9,\ee where $\eta_q=\pm
1$. Since Eq.(\ref{exa3}) is not satisfied, $\eta_{3,5,7,9}$ must
not simultaneously be $-1$. Now consider a new $C$ \be C'\equiv
{1\over 2}\Big[C-\eta_3 C_3-\eta_5 C_5-\eta_7 C_7-\eta_9
C_9\Big].\ee It is obvious that upon
$\{\alpha_{k1},\alpha_{k2},\alpha_{k3}\}\leftrightarrow
\{\alpha_{l1},\alpha_{l2},\alpha_{l3}\}$ $C'\rightarrow -C'$.
Moreover by construction $C'$ only contains those $C_q$ whose
$\eta_q=-1$. Now use $C'$ as the starting $C$ and repeat the above
operation until we reach a final $C'$ for which Eq.(\ref{exa3})
holds for all triplet exchanges and for all $(i,j)$. Since at each
stage of obtaining $C'$ certain $C_q$ are projected out, there
must be missing $q$ components in the final $C$.  However we have
already proven that any $C$ that satisfy Eq.(\ref{exa3}) for all
$(i,j)$ pair must satisfy \Eq{c2}. However \Eq{c2} contains all
four $q$ components for all pair $(i,j)$. Consequently we have
reached a contradiction. Therefore it must be possible to make
Eq.(\ref{exa3}) hold true for all pairs $(i,j)$ for any ground
state solution satisfying \Eq{su34}.

Although we have chosen $p=3$ and $m=3$ in the above discussion, it
should be clear that our proof can be generalized to any odd $p$ ,
any $m$. Thus we have proven that Eq.(\ref{su3wf}) is the unique
ground state of Eq.(\ref{su31}).

\begin{figure}[htb]
\centerline{
   \includegraphics[angle=0,scale=0.35]{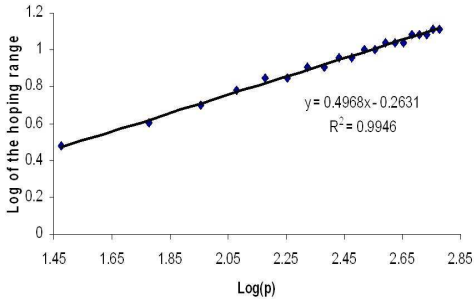}}
\caption{A log-log plot of the hopping range and $p$ from 30 to
600. The straight line is the best fit.  The vertical axis is the
log of the hopping range and the horizontal one is $\log p$. The
hopping range scales as $p^{\frac{1}{2}}$ } \label{Fig:HopRangSu3}
\end{figure}

It is straightforward to prove that Eq.(\ref{su3}) is a
center-of-mass position conserving pair hopping model, i.e.,
\begin{eqnarray}
H&&=\kappa
\sum_{j,L,L_3}\sum_{l,l_3}\sum_{k,k_3}F^{j,L,L_3}_{l,l_3}F^{j,L,L_3}_{k,k_3}\nonumber\\&&
c^\dag_{(l,l_3)}c^\dag_{(j-l+\frac{1}{2},L_3-l_3)}
c_{(j-k+\frac{1}{2},L_3-k_3)}c_{(k,k_3)}.\label{php}
\end{eqnarray}
 The central
steps are 1) viewing the weight space of $(mp,0)$ as a triangular
lattice, and 2) decomposing the two spin states in \Eq{su3} as
linear combination of products of single spin state. Due to space
limitation, the result will be published elsewhere. Using the
explicit expression for the $F$'s in \Eq{php} (lengthy hence is
omitted here) we can estimate the hopping range. In
Fig.(\ref{Fig:HopRangSu3}) we plot the hopping range versus $p$
for $p=30\ra p=600$. While the linear dimension of the lattice
scales as $p$, the hopping range scales as $p^{\frac{1}{2}}$,
hence the hopping is long-ranged as the SU(2) case.

Finally, we demonstrate the presence of an excitation gap within
the single mode approximation (SMA)\cite{girvin1986}. Analogous to
the SU(2) case it is possible to view the $(k,0)$ SU(3) multiplet
as the ``lowest Landau level''(LLL) of a particle running on
CP$^2$ under the action of a U(1) background magnetic
field\cite{karabali2002}. If we parameterize the fundamental SU(3)
spinor as $\left(1,z_1,z_2\right)/\sqrt{1+|z_1|^2+|z_2|^2},$ where
$z_i=x_{i}+i y_{i}$, and take the flat-space limit (i.e., restrict
$|z_{1,2}|<<1$) the single-particle orbitals in the LLL become
\begin{eqnarray}
\Phi_{l_1,l_2}(z_1,z_2)=\frac{1}{\sqrt{4\pi^22^{l_1}2^{l_2}l_1!l_2!}}
z_1^{l_1}z_2^{l_2}e^{-(|z_1|^2+|z_2|^2)/4},\nonumber
\label{sp_state}
\end{eqnarray}
where $l_{1,2}$ are non-negative integers. We recognize that the
above result is the product of two LLL wavefunctions in two space
dimensions. Thus in the flat-space limit, the LLL in CP$^2$
becomes the direct product of the LLLs in two quantum Hall planes.
This reduction allows us to perform the SMA calculation pretty
much in parallel to that for the ordinary quantum Hall
effect.\cite{girvin1985,girvin1986} in the following we summarize
the results. Within SMA the excitation energy is given by
\begin{eqnarray}
\Delta(\v k)=f(\v k)/s(\v k).
\end{eqnarray}
Here $\v k=(k_1, k_2)$ where $k_{1,2}$ are the complex wave
vectors associated with the two quantum Hall planes, and $f(\v k)$
and $s(\v k)$ are given by \be &&f(\v
k)=(1/N)<\Psi_m|[\rho^\dagger_{\v k},[V,\rho_{\v k}]]|\Psi_m> \nn
&&s(\v k)=(1/N)<\Psi_m|\rho^\dagger_{\v k}\rho_{\v k}|\Psi_m>. \ee
In the above equation $\rho_{\v k}$ and $V$ are the the density
operator and the inter-particle potential projected onto the LLL.
Straightforward calculation gives,
\begin{eqnarray}
&\!&f(\v k)\!=\!\frac{1}{2}\sum_{\v q}v(|\v q|)(e^{\frac{\bar{\v
q}\v k}{2}}\!-\!e^{\frac{\bar{\v k}\v q}{2}}) [s(\v
q)e^{-\frac{|\v k|^2}{2}}(e^{-\frac{\bar{\v k}\v
q}{2}}\!-\!e^{-\frac{\bar{\v q}\v k}{2}})\nonumber\\&\!&+s(\v k+\v
q)(e^{\frac{\bar{\v k}\v q}{2}}-e^{\frac{\bar{\v q}\v k}{2}})],
\label{f}
\end{eqnarray}
where $v(|\v q|)$ is the Fourier transformation of the potential,
which is required to be positive indicating the repulsive
interaction to ensure the excitation energy to be positive.   On
the other hand, $s(\v k)$ can be related to the radial
distribution function $g(\vec{r})$ by
\begin{eqnarray}
s(\v k)=e^{-\frac{|\v k|^2}{2}}\!+\!\rho\int
d^4re^{-i\vec{k}\cdot\vec{r}}[g(\vec{r})\!-\!1]\!+\!\rho(2\pi)^4\delta^4(\vec{k})
\label{s}
\end{eqnarray}
where $\rho$ is the average density and $\vec{k}$=(Re($k_1$),
Im($k_1$), Re($k_2$), Im($k_2$)).  After some algebra, for small
$|\v k|$ it can be shown that $\Delta(\v
k)=(a|k_1|^4+b|k_1|^2|k_2|^2+a|k_2|^4)/(
c|k_1|^4+d|k_1|^2|k_2|^2+c|k_2|^4)$, which remains finite as $\v
k$ approaches to zero in any direction.

In summary, we have constructed a two dimensional center-of-mass
conserving pair hopping model which exhibit incompressible quantum
liquid ground state. Although throughout the Letter we have
focused on SU(3) whose weight space is two dimensional, our
construction can easily be generalized to SU(N) giving rise to
models for incompressible quantum liquid in higher dimensions.
Particularly, the SU(4) model can be applied to the
four-dimensional quantum Hall effect proposed by Zhang and Hu
\cite{Zhang2001Science, Chern2006, bernevig2002ANN}.

We deeply appreciate the discussion with Darwin Chang.  CHC is
supported by ERATO-SSS, Japan Science and Technology Agency. DHL
is supported by the Directior, Office of Science, Office of Basic
Energy Sciences, Materials Sciences and Engineering Division, of
the U.S. Department of Energy under Contract No.
DE-AC02-05CH11231.

%\bibliographystyle{unsrt}
%\bibliography{FMIPRL}

\end{document}